\documentclass[english,11pt,aps,prd,a4paper,preprintnumbers,nofootinbib,showpacs,superscriptaddress, notitlepage]{revtex4-2} 
\newcommand{\AddrUNAM}{Instituto de F\'isica, Universidad Nacional Aut\'onoma de M\'exico, A.P. 20-364, Ciudad de M\'exico 01000, M\'exico.}
\usepackage{color}
\usepackage{float}
\usepackage{graphicx}
\usepackage{multirow}
\usepackage{amsmath}
\usepackage{amssymb}
\usepackage[colorlinks=true,citecolor=darkred,urlcolor=darkred, pdfborder={0 0 0}]{hyperref}
\usepackage[normalem]{ulem}
\definecolor{darkred}{rgb}{0.6,0,0}
\definecolor{drkgrn}{RGB}{0, 51, 0}
\usepackage{soul}

\newcommand{\U}[1]{\, \mathrm{#1}}

\definecolor{linkcolor}{rgb}{0,0,0.5}

\def\gsim{\raise0.3ex\hbox{$\;>$\kern-0.75em\raise-1.1ex\hbox{$\sim\;$}}}
\def\lsim{\raise0.3ex\hbox{$\;<$\kern-0.75em\raise-1.1ex\hbox{$\sim\;$}}}

\def\beqn#1{\begin{equation}\label{#1}}
\def\eeqn{\end{equation}}

\def\beqa#1{\begin{eqnarray}\label{#1}}
\def\eeqa{\end{eqnarray}}

\def\Z2{$\mathcal{Z_2}$}

\newcommand {\ignore}[1]{}

\begin{document}

\title{\boldmath Flavored axion in the UV-complete Froggatt-Nielsen models}
\author{Leon M.G. de la Vega}\email{leonm@estudiantes.fisica.unam.mx}\affiliation{\AddrUNAM}
\author{Newton Nath}\email{newton@fisica.unam.mx}\affiliation{\AddrUNAM}
\author{Stefan Nellen} \email{stefannellen@estudiantes.fisica.unam.mx}\affiliation{\AddrUNAM}
\author{Eduardo Peinado} \email{epeinado@fisica.unam.mx}\affiliation{\AddrUNAM}

\vspace{2.0cm}
\begin{abstract}
{\noindent We propose UV-completions of Froggatt-Nielsen-Peccei-Quinn models of fermion masses and mixings with flavored axions, by incorporating heavy fields.
Here, the $U(1)$ Froggatt-Nielsen symmetry is identified with the Peccei-Quinn symmetry to solve the strong CP problem along with the mass hierarchies of the Standard Model fermions. We take into account  leading order contributions to the fermion mass matrices giving rise to Nearest-Neighbour-Interaction structure in the quark sector and $A_2$ texture in the neutrino sector. A comprehensive numerical analysis has been performed for the fermion mass matrices. Subsequently,  we investigate the resulting axion flavor violating couplings and the axion-photon coupling arising from the model. 
}
\end{abstract}
%

\maketitle

\section{Introduction}
The axion~\cite{Peccei:1977hh,Wilczek:1977pj,Weinberg:1977ma} is one of the best motivated elementary particles beyond the Standard Model (BSM).
It provides the most cogent  solution to the ``strong CP problem" 
\footnote{It is known that the theory of strong interactions i.e., QCD violates CP symmetry. However, the smallness of CP violation in QCD has been a long-standing puzzle in particle physics, called as the ``strong CP problem". At present, the most stringent bounds on the CP-violating term come  from experimental limits on the neutron electric dipole moment and result in $ \lesssim 10^{-10} $~\cite{Baker:2006ts,Zyla:2020zbs}.}
 of quantum chromodynamics  (QCD), where the spontaneous breaking of a global $U(1)$ symmetry, called the Peccei-Quinn (PQ) symmetry $U(1)_{PQ}$, gives rise to a pseudo-Nambu-Goldstone (pNG) boson~\cite{Wilczek:1977pj,Weinberg:1977ma}.
It is also a good candidate for cold dark matter within a certain allowed parameter space of the axion scale~\cite{Dine:1982ah,Abbott:1982af,Preskill:1982cy,Davis:1986xc,Kim:2008hd}.
In the literature, there exist two common approaches  to realize the PQ symmetry that give rise to the axion, in one case the Higgs sector is extended (called the DFSZ model) \cite{Dine:1981rt,Zhitnitsky:1980tq}, whereas in a different scenario heavy quarks are introduced (called the KSVZ model)~\cite{Kim:1979if,Shifman:1979if}.
As per the experimental testability of the axion is concerned, it is the axion couplings to the photon $ g_{a\gamma} $ and the electron $ g_{a e} $ that play very important roles. One can establish a relation between these couplings with the PQ charges of fermions~\cite{Srednicki:1985xd}.
It is the constant $E/N$, i.e., the ratio of the electromagnetic over the QCD anomaly coefficient which determines the strength of the photon-axion coupling, and it is a free parameter in generic axion models.

Another puzzle in the Standard Model (SM) is the unexplained hierarchies in masses and mixings between elementary particles or the ``hierarchical flavor structure" of the Yukawa couplings. Many flavor models have been studied to address the SM mass hierarchies involving additional symmetries.
Among them, the Froggatt-Nielsen (FN) symmetry emerges as a leading candidate to account for the 
flavor hierarchies~\cite{Froggatt:1978nt}. In this mechanism, one introduces a new complex scalar field called flavon, whose vacuum expectation value ($ vev $) generates the hierarchical flavor structure, where a global flavor symmetry $U(1)_{FN}$ is imposed.
Now, here one can ask, if it is possible to use the PQ symmetry as a flavor symmetry in order to explain the hierarchical flavor structure while solving the strong CP problem of the SM. Indeed, in~\cite{PhysRevLett.48.11,Wilczek:1982rv,Davidson:1983fy} a connection between the PQ symmetry with flavor symmetries has been addressed. 
Recently, many attempts have been made where the PQ symmetry $U(1)_{PQ}$ is unified with the FN symmetry  $U(1)_{FN}$ to address both issues of the SM. The resultant axion that arises from this flavor-dependent framework has been called ``flaxion"~\cite{Ema:2016ops,Carone:2020nlx}, ``axiflavon"~\cite{Calibbi:2016hwq,Arias-Aragon:2017eww,Linster:2018avp}, or ``flavorful axion"~\cite{Bjorkeroth:2017tsz,Bjorkeroth:2018dzu,Bonnefoy:2019lsn}.

In this work, we study various low energy phenomenologies of the flavor-dependent axion that arises from the PQ symmetry providing a solution to the strong CP problem. Here, only the third family of quarks is generated at dimension four while other terms are introduced at higher dimensions. We also extended the flavor structure to the leptonic sector.
In doing so, we identify the $U(1)_{PQ}$ symmetry with the $U(1)_{FN}$ that gives rise to the fermion hierarchies, where the axion field $ \sigma $ is treated like a flavon. We consider the DFSZ type axion models and show that one can construct a Nearest-Neighbour-Interaction (NNI) structure for both the up- and down-quark mass matrices
\begin{equation}\label{eq:NNI}
M_{u/d} = \left( \begin{array}{ccc}
0 & \times & 0 \\
\times & 0 & \times \\
0 & \times & \times
\end{array}\right)\;,
\end{equation}
which was originally proposed in~\cite{Branco:1988iq}.
For the DFSZ models, we affix the PQ charges to all the SM fermions as well as to the Higgs sector in such a way that after the PQ and electroweak (EW) symmetry breaking, one ends up with the NNI structure of the quark mass matrices as given by Eq.~\ref{eq:NNI}. 
%
%
Later, we present the UV-complete Froggatt-Nielsen-Peccei-Quinn (FNPQ) models  by incorporating heavy fermionic fields~\footnote{Notice that  UV-completion of such flavored axion  within the formalism of SUSY models has been discussed in~\cite{Bonnefoy:2019lsn}, where authors have added two singlet superfields. Also, in their formalism they have considered the gauged FN symmetry in absence of any heavy vector-like or chiral fields.}.
Two examples have been adopted within the framework of the type-I and -II Dirac seesaw mechanisms.
In both the scenarios, it is shown that once we integrate out heavy fields,
one finds NNI-type quark mass matrices.

Also, to explain the masses and mixings of the leptonic sector, we find the diagonal charged-lepton mass matrix in the FNPQ-framework. Moreover, within the framework of the type-I seesaw mechanism~\cite{Minkowski:1977sc,Yanagida:1979as,GellMann:1980vs,Mohapatra:1979ia,Schechter:1980gr},
 we obtain an $ A_2 $-type~\cite{Frampton:2002yf} neutrino mass matrix which is allowed by the latest global analysis of the neutrino oscillation data~\cite{Capozzi:2017ipn,deSalas:2020pgw,Esteban:2020cvm}.
We also perform a numerical analysis of the quarks as well as leptonic mass matrices to find masses and mixings of fermions. Later, we explore flavor violating decays with axions, where we estimate the branching ratios (BR) of $ K^+ \rightarrow \pi^+ a  $ and $ B^+ \rightarrow K^+ a  $. These further help us to constrain the axion breaking scale.
Furthermore, because of the presence of two Higgs doublets in the DFSZ axion model, we calculate the
flavor violating process $ t \rightarrow h c $. Finally, the cosmological consequences of the FNPQ axion are discussed.

We plan the manuscript as below. In Sec.~\ref{sec:FN-NNI} we give a brief overview of the FN symmetry along with the NNI quark mass matrices. A detailed discussion of our theoretical framework that leads to NNI quark textures together with the UV-complete mechanisms are presented in Sec.~\ref{sec:Framework} and its subsequent sub-sections, while the leptonic sector is described in Sec.~\ref{sec:Lepton}. The scalar sector and axion couplings are presented in Sec.~\ref{sec:ScalarSector}, and Sec.~\ref{sec:AxionCoupling}, respectively. We analyze our numerical results in Sec.~\ref{sec:results}, and conclusions are summarized in Sec.~\ref{sec:conclusion}. 


\section{$U(1)_{FN}$ and Nearest-Neighbour-Interaction}\label{sec:FN-NNI}
Here, we start with a brief overview of the Froggatt-Nielsen symmetry and Nearest-Neighbour-Interaction structure for the fermion mass matrices.
In order to address the hierarchical structure of the fermion masses Froggatt and Nielsen developed a symmetry called FN-symmetry~\cite{Froggatt:1978nt}, where they introduced a SM singlet ``flavon" field $ \sigma $. In their framework, one can express the necessary Yukawa terms for the quark sector as
\begin{equation}\label{eq:FN}
-\mathcal{L} \supset y_{ij}^d\left( \frac{\sigma}{\Lambda} \right)^{n_{ij}^d} \overline Q_i H d_{Rj}
	+  y_{ij}^u\left( \frac{\sigma}{\Lambda} \right)^{n_{ij}^u} \overline Q_i \widetilde H u_{Rj} \;,
\end{equation}
where $ n_{ij}^{u/d}$ are complex numbers, $ \sigma $ is the flavon field and $ \Lambda $ is the scale of flavor dynamics. Also, $Q_i$, $u_{Ri}$, $d_{Ri}$ for $ i,j = 1, 2, 3 $ represent the left-handed quark doublet, right-handed up-type quark, and
right-handed down-type quark, respectively. Notice that one can write similar terms for the leptonic sector.
Once, the singlet acquires its $ vev $ and breaks the symmetry, it leads to hierarchy in the masses of quarks and leptons. Thus, one can notice from Eq.~\ref{eq:FN} that the hierarchy of the Yukawa couplings can be correlated to the smallness of $ \langle\sigma\rangle/\Lambda $, and hence to the fermion mass matrices $ m_{ij} = y_{ij} v (\langle\sigma\rangle/\Lambda)^{n} $, with $ v $ being the SM Higgs vacuum expectation  value.

Furthermore, this global $U(1)$ symmetry can be used to obtain NNI-textures in the quark sector. The NNI-texture for a $N\times N$ mass matrix consists of a general matrix $m$ with $m_{ij}\neq 0$ only for $i = j \pm 1$ or $i=N=j$.
Notice that one can connect the NNI-texture with the Fritzsch texture~\cite{Fritzsch:1977vd,Fritzsch:1979zq} with an additional constraint of Hermiticity on both the quark mass matrices.  However, the Fritzsch texture was ruled out as it failed to simultaneously generate the small value of the CKM matrix element  $V_{cb}$ together with the large value of top-quark mass $ m_t $~\cite{Ishimori:2010au,Fritzsch:2011cu}. The NNI texture is phenomenologically compatible with the quark sector, but contains more degrees of freedom in contrast to the experimental constraints, i.e. 6 masses, 3 angles and a CP-violating phase of the quark sector.
In the SM, the NNI texture may be obtained through Weak-Basis (WB) transformations from an arbitrary Yukawa matrix \cite{Branco:1988iq}, and thus imposes no physical constraints. However, in models with more than one scalar a global symmetry that imposes the NNI on the quark mass matrix imposes a particular structure of scalar interactions \cite{Branco:2010tx}, potentially leading to physical observables. It is well known that the NNI structure can reproduce the CKM mixing pattern, while demanding a hierarchy between its elements to obtain that of the quark masses \cite{Harayama:1996jr}. One particular feature is that the (3, 3) element is of order $m_t$ ($m_b$) in the up (down) sector, while the other entries are heavily suppressed. The proposed framework for the flavored PQ symmetry accommodates the DFSZ style of axion models, while leading to the NNI structure of the quark mass matrices, explaining naturally the largeness of the (3, 3) element, and the $A_2$ structure of the neutrino mass matrix in the diagonal charged lepton basis.
\section{Fermion masses in flavored DFSZ axion models}\label{sec:Framework}
Now, we propose UV-complete models of quark masses, where a $U(1)$ global symmetry enforces NNI texture in the up and down quark mass matrices. The $(3,3)$ entry of the matrix is generated by a dimension-4 operator, while the other entries are obtained by dimension-5 operators with the addition of a symmetry-breaking flavon, thus obtaining mass hierarchies in a Froggatt-Nielsen style. Additionally, through the color anomalies of the global $U(1)$ symmetry we obtain a QCD axion from the flavon fields, which solves the strong CP problem and is a Dark Matter candidate. We present two realizations within the framework  of the DFSZ style of QCD axions.

In the DFSZ axion model, two Higgs doublets are needed in addition to the scalar singlet which introduces the PQ symmetry breaking scale.
We list the relevant field contents with their corresponding $SU(2)_L\times U(1)_Y$ and $U(1)_{PQ}$  charges in Table~\ref{tab:DFSZ1}. We have derived the PQ charges by imposing the appearance of the NNI structure at dimension-5, while choosing opposite charges for left and right handed fields in the spirit of simplicity. Additionally we have followed the convention followed in DFSZ models where the charges of the Higgs doublets are normalized to unity.
 \begin{center}
\begin{table}[h] 
\scriptsize
\begin{tabular}{ c | c c c | c c c }
\hline
Fields/Symmetry  & \hspace{1cm}  $Q_{iL} $ \hspace{1cm} & \hspace{1cm}   $u_{iR} $ \hspace{1cm} &\hspace{1cm}  $d_{iR}$ \hspace{1cm} & $H_u$ &  $H_d$  & $\sigma$  \\
\hline
\hline
$SU(2)_L\times U(1)_Y$   &  (2, 1/6)  &  (1, 2/3) & (1, -1/3)  & (2, -1/2) & (2, 1/2) & (1, 0) \\  
\hline
$U(1)_{PQ}$  & (9/2, -5/2,  1/2) &  (-9/2, 5/2,  -1/2) &  (-9/2, 5/2, -1/2) & 1 & 1 & 1 \\
\hline
  \end{tabular}
\caption{\footnotesize Field content and transformation properties of the PQ-symmetry under the DFSZ type-I seesaw model, where $i = 1, 2, 3$ represent families of three quarks.}
\label{tab:DFSZ1}
\end{table}
\end{center}
The effective Lagrangian that describes the up-quark sector in the model with the $ SM\times U(1)_{PQ}$ charges of Table \ref{tab:DFSZ1} can be written as
\begin{align}\label{eq:type-1up}
\mathcal{L}  & \supset \frac{C^{u}_{11}}{\Lambda^8} \overline{Q}_{1L} H_u u_{1R} \sigma^8 + \frac{C^{u}_{12}}{\Lambda} \overline{Q}_{1L} H_u u_{2R} \sigma  + \frac{C^{u}_{13}}{\Lambda^4} \overline{Q}_{1L} H_u u_{3R} \sigma^4 + \frac{C^{u}_{21}}{\Lambda} \overline{Q}_{2L} H_u u_{1R} \sigma +\frac{C^{u}_{22}}{\Lambda^4} \overline{Q}_{2L} \widetilde{H}_d u_{2R} \sigma^{*4}\nonumber \\
& +  \frac{C^{u}_{23}}{\Lambda} \overline{Q}_{2L} \widetilde{H}_d u_{3R} \sigma^*  + \frac{C^{u}_{31}}{\Lambda^4} \overline{Q}_{3L} H_u u_{1R} \sigma^4 + \frac{C^{u}_{32}}{\Lambda} \overline{Q}_{3L} \widetilde{H}_d u_{2R} \sigma^*  
  + y^u_{33} \overline{Q}_{3L} H_u u_{3R}    \;,
\end{align}
where $ C^{u}_{ij} $ represents coupling constant and $ \Lambda $ is the cut-off scale of the model.
Similarly, one can write the Lagrangian for the down-quark sector
\begin{align}\label{eq:type-1down}
\mathcal{L}  & \supset \frac{C^{d}_{11}}{\Lambda^8} \overline{Q}_{1L} H_d d_{1R} \sigma^8 + \frac{C^{d}_{12}}{\Lambda} \overline{Q}_{1L} H_d d_{2R} \sigma + \frac{C^{d}_{13}}{\Lambda^4} \overline{Q}_{1L} H_d d_{3R} \sigma^4 + \frac{C^{d}_{21}}{\Lambda} \overline{Q}_{2L} H_d d_{1R} \sigma 
 +\frac{C^{d}_{22}}{\Lambda^4} \overline{Q}_{2L} \widetilde{H}_u d_{2R} \sigma^{*4}  \nonumber \\
& +    \frac{C^{d}_{23}}{\Lambda} \overline{Q}_{2L} \widetilde{H}_u d_{3R} \sigma^*
 + \frac{C^{d}_{31}}{\Lambda^4} \overline{Q}_{3L} H_d d_{1R} \sigma^4
 + \frac{C^{d}_{32}}{\Lambda} \overline{Q}_{3L} \widetilde{H}_u d_{2R} \sigma^* 
 + y^d_{33} \overline{Q}_{3L} H_d d_{3R} \;,
\end{align}
with a similar meaning for $C_{ij}^d$ and $\Lambda$ as the up-quark sector. Notice that these Higgs doublets couple to both the quark sectors.
After the symmetry breaking, and counting terms up to dimension-7, we find  NNI-type quark-mass matrices as
\footnote{
Notice that the higher dimensional operators are either suppressed by a high enough energy scale compared to dimension-4 or 5, or one can forbid them  by using an additional symmetry, and hence one can safely deal with the NNI-type quark texture.
}
\begin{equation}\label{eq:MassQuark}
M_{u/d} = \left( \begin{array}{ccc}
0 & \varepsilon v_{u/d} C_{12}^{u/d}   & 0 \\
\varepsilon v_{u/d} C_{21}^{u/d}   & 0 & \varepsilon v_{d/u}  C_{23}^{u/d}  \\
0  &\varepsilon v_{d/u} C_{32}^{u/d}   & y_{33}^{u/d} v_{u/d}
\end{array}\right)\;,
\end{equation}
where, $ \varepsilon = \langle \sigma\rangle/ \Lambda $ or $\langle \sigma\rangle^*/ \Lambda $. For typical values of   $\varepsilon\sim 0.2 $, one can safely neglect terms proportional to $ \varepsilon^{4} $ and $ \varepsilon^{8} $ as has been pointed in Eqs.~\ref{eq:type-1up}, \ref{eq:type-1down}. In this framework the mass hierarchy between the third family and the first two is naturally explained by the fact that the (3, 3) entry is generated at dimension-4 and the rest of the entries are generated at dimension-5 . Additionally, one can explain the hierarchy between $m_t$ and $m_b$ by a hierarchy between $v_u$ and $v_d$, as in two Higgs doublet models with natural flavor conservation.


\subsection{The UV-completion: DFSZ type-I Seesaw}\label{sec:DFSZ-I}
First, we consider a  type-I Dirac seesaw model, where the heavy mediators are vector-like quarks. Furthermore, this restriction will exclude them from contributing to the anomalous couplings of the axion.
The UV-complete operator for the up-quark sector, as given by Eq.~\ref{eq:type-1up}, within the DFSZ type-I seesaw formalism can be achieved as follows
\begin{align}\label{eq:uv-1}
\mathcal{L}^{UV}_{u} &  \supset  \mathcal{Y}^{u}_{12} \overline{Q}_{1L} H_u F^{12}_{uR} + \mathcal{M}^{u}_{12} \overline{F^{12}_{uR}} F^{12}_{uL} + \mathcal{Y^\prime}^{u}_{12}~ \overline{F^{12}_{uL}} \sigma u_{2R} \nonumber \\
&  + \mathcal{Y}^{u}_{21} \overline{Q}_{2L} H_u F^{21}_{uR} + \mathcal{M}^{u}_{21} \overline{F^{21}_{uR}} F^{21}_{uL} + \mathcal{Y^\prime}^{u}_{21} \overline{F^{21}_{uL}} \sigma u_{1R} \nonumber \\
& + \mathcal{Y}^{u}_{23} \overline{Q}_{2L} \widetilde{H}_d F^{23}_{uR} + \mathcal{M}^{u}_{23} \overline{F^{23}_{uR}} F^{23}_{uL} + \mathcal{Y^\prime}^{u}_{23}~ \overline{F^{23}_{uL}} \sigma^* u_{3R} \nonumber \\
&  + \mathcal{Y}^{u}_{32} \overline{Q}_{3L} \widetilde{H}_d F^{32}_{uR} + \mathcal{M}^{u}_{32} \overline{F^{32}_{uR}} F^{32}_{uL} + \mathcal{Y^\prime}^{u}_{32} \overline{F^{32}_{uL}} \sigma^* u_{2R}
 \;,
\end{align}
where  $ F^{ij}_{qC} $  are the vector like fields and their PQ-charges are given in Table~\ref{tab:DFSZ2}.
 
\begin{figure}
\centering
\includegraphics[width=0.5\textwidth]{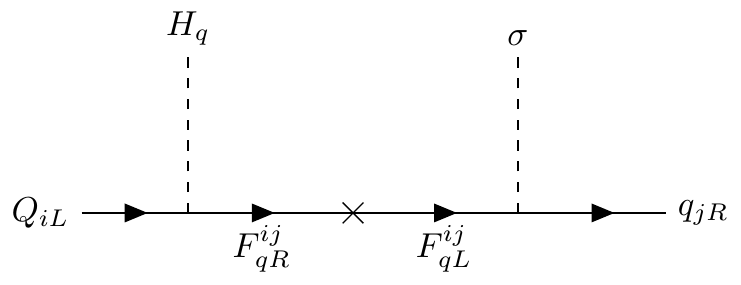}
\caption{\footnotesize UV complete diagram within the DFSZ type-I seesaw framework as apparent from Eqs.~\ref{eq:uv-1} and \ref{eq:uv-2}.
}
\label{fig:uvdiagramq}
\end{figure}

Similarly, the UV-completion of the down-quark sector as given by  Eq.~\ref{eq:type-1down}  can be achieved by using $ F^{ij}_{qC} $ (see Table~\ref{tab:DFSZ2} for their charges) and the corresponding Lagrangian can be written as
\begin{align}\label{eq:uv-2}
\mathcal{L}^{UV}_{d} &  \supset  \mathcal{Y}^{d}_{12} \overline{Q}_{1L} H_d F^{12}_{dR} + \mathcal{M}^{d}_{12} \overline{F^{12}_{dR}} F^{12}_{dL} + \mathcal{Y^\prime}^{d}_{12}~ \overline{F^{12}_{dL}} \sigma d_{2R} \nonumber \\
 & +  \mathcal{Y}^{d}_{21} \overline{Q}_{2L} H_d F^{21}_{dR} + \mathcal{M}^{d}_{21} \overline{F^{21}_{dR}} F^{21}_{dL} + \mathcal{Y^\prime}^{d}_{21} \overline{F^{21}_{dL}} \sigma d_{1R} \nonumber \\
& + \mathcal{Y}^{d}_{23} \overline{Q}_{2L} \widetilde{H}_u F^{23}_{dR} + \mathcal{M}^{d}_{23} \overline{F^{23}_{dR}} F^{23}_{dL} + \mathcal{Y^\prime}^{d}_{23}~ \overline{F^{23}_{dL}} \sigma^* d_{3R} \nonumber \\
 & +  \mathcal{Y}^{d}_{32} \overline{Q}_{3L} \widetilde{H}_u F^{32}_{dR} + \mathcal{M}^{d}_{32} \overline{F^{32}_{dR}} F^{32}_{dL} + \mathcal{Y^\prime}^{d}_{32} \overline{F^{32}_{dL}} \sigma^* d_{2R}
 \;.
\end{align}
In Fig.~\ref{fig:uvdiagramq} we show the Feynman diagram corresponding to Eqs.~\ref{eq:uv-1} and \ref{eq:uv-2}.
%
 \begin{center}
\begin{table}[t] 
\scriptsize
\begin{tabular}{c | c c c c c c c c  } 
\hline
Fields/Symmetry  & \hspace{1mm} $F^{12}_{uC} $ \hspace{1mm}  &  \hspace{1mm} $F^{21}_{uC} $ \hspace{1mm} &   \hspace{1mm}   $F^{23}_{uC} $ \hspace{1mm} &   \hspace{1mm} $F^{32}_{uC} $ \hspace{1mm} & \hspace{1mm} $F^{12}_{dC} $ \hspace{1mm}  &  \hspace{1mm} $F^{21}_{dC} $ \hspace{1mm} &   \hspace{1mm}   $F^{23}_{dC} $ \hspace{1mm} &   \hspace{1mm} $F^{32}_{dC} $ \hspace{1mm}   \\
\hline \hline
$U(1)_Y$   &  2/3  & 2/3   & 2/3 & 2/3 & -1/3 &  -1/3 & -1/3 & -1/3 \\
\hline
$U(1)_{PQ}$    &  7/2   & -7/2  & -3/2 & 3/2 & 7/2 &  -7/2 & -3/2 & 3/2  \\
\hline
  \end{tabular}
\caption{\footnotesize Vector like fermions and their transformation properties of the PQ-symmetry under the DFSZ type-I seesaw model, where $C = L, R$.}\label{tab:DFSZ2}
\end{table}
\end{center}
%
%
%
%

After the PQ and EW symmetry breaking by the $ vev $s $v_{\sigma}$, $\langle H_u \rangle$ and $\langle H_d \rangle$ we obtain the following quark mass matrices in the $(\overline{Q}_L,F_{(u/d)L}) \times (q_R,F_{(u/d)R})$ basis as
\begin{equation}\label{eq:seesawmatrix}
M_{u/d} = \begin{pmatrix}
M_{Q_{L}q_{R}} & M_{Q_{L}F_{R}} \\ 
M_{F_{L}q_{R}} & M_{F_{L}F_{R}}
\end{pmatrix}_{7\times7} ,
\end{equation}
where the 4 submatrices of $ M_{u/d} $  are given by

\begin{center}
\begin{eqnarray}
M_{Q_{L}q_{R}} &=&\begin{pmatrix}
0 & 0 & 0 \\ 
0 & 0 & 0 \\ 
0 & 0 & y_{33}^{u/d} v_{u/d}
\end{pmatrix} ,\\
M_{Q_{L}F_{R}} &=&\begin{pmatrix}
\mathcal{Y}^{u/d}_{12} v_{u/d} & 0 & 0 & 0 \\ 
0 & \mathcal{Y}^{u/d}_{21} v_{u/d} & \mathcal{Y}^{u/d}_{23} v_{d/u} & 0 \\ 
0 & 0 & 0 & \mathcal{Y}^{u/d}_{32} v_{d/u}
\end{pmatrix} ,\\
M_{F_{L}q_{R}} &=&\begin{pmatrix}
0 & \mathcal{Y^\prime}^{u/d}_{12} v_{\sigma} & 0  \\ 
\mathcal{Y^\prime}^{u/d}_{21} v_{\sigma} & 0 & 0 \\ 
0 & 0 & \mathcal{Y^\prime}^{u/d}_{23} v_{\sigma}^*  \\
0 & \mathcal{Y^\prime}^{u/d}_{32} v_{\sigma}^* & 0
\end{pmatrix} ,\\
M_{F_{L}F_{R}} &=& \mathrm{diag}(\mathcal{M}_{12}^{u/d},\mathcal{M}_{21}^{u/d},\mathcal{M}_{23}^{u/d},\mathcal{M}_{32}^{u/d})  \;.
\end{eqnarray}
\end{center}

Now, the light quark mass matrix in this Dirac type-I seesaw scenario can be written as
\begin{equation}\label{eq:seesawformula}
m_{u/d} = M_{Q_{L}q_{R}} - M_{Q_{L}F_{R}} M^{-1}_{F_{L}F_{R}} M_{F_{L}q_{R}} \; ,
\end{equation}
at leading order. The resulting form of this matrix is given by
\begin{align}
m_{u/d} & = \begin{pmatrix}
0 & \frac{ \mathcal{Y}^{u/d}_{12}  \mathcal{Y^\prime}^{u/d}_{12}}{\mathcal{M}_{12}} v_{u/d} v_{\sigma}   & 0 \\ 
\frac{\mathcal{Y}^{u/d}_{21}  \mathcal{Y^\prime}^{u/d}_{21}}{\mathcal{M}_{21}} v_{u/d} v_{\sigma}   & 0 & \frac{\mathcal{Y}^{u/d}_{23}  \mathcal{Y^\prime}^{u/d}_{23}}{\mathcal{M}_{23}} v^*_{d/u} v^*_{\sigma}   \\ 
0 & \frac{\mathcal{Y}^{u/d}_{32}  \mathcal{Y^\prime}^{u/d}_{32}}{\mathcal{M}_{32}} v^*_{d/u} v^*_{\sigma}  & y_{33}^{u/d} v_{u/d}
\end{pmatrix}\;,  \nonumber \\
%
%
& = \begin{pmatrix}
0 & \mathbf{A}_{u/d} & 0 \\
\mathbf{B}_{u/d} & 0 & \mathbf{C}_{u/d} \\
0 & \mathbf{D}_{u/d} & \mathbf{E}_{u/d}
\end{pmatrix} \;.
\label{eq:massqf}
\end{align}
where $\mathbf{A}, \, \mathbf{B}, \, \mathbf{C}, \, \mathbf{D}$, and $\mathbf{E}$ are complex entries. \\

\subsection{The UV-completion: DFSZ type-II Seesaw}
%
%
%
As an alternative, we consider a type-II Dirac seesaw framework of a DFSZ model.
%
Within this model, the UV-completion of the quark-sectors (see Eqs.~\ref{eq:type-1up}, \ref{eq:type-1down}) can be achieved by introducing two additional BSM doublets namely, $ \Phi_{u} (2, -1/2,2) $, and $ \Phi_{d} (2,1/2, 2) $, where parentheses in the brackets signify $ SU(2)_L \times U(1)_{Y}\times U(1)_{PQ} $, respectively.
Here, no additional heavy quark states are added.
 We express the UV-complete Lagrangian for the up-quark sector as follows:
\begin{align}\label{eq:uv-3}
\mathcal{L}^{UV}_{u} &  \supset  \mathcal{Y}^{u}_{12} \overline{Q}_{1L} \Phi_u u_{2R} + \mathcal{Y}^{u}_{21} \overline{Q}_{2L} \Phi_u u_{1R} + \kappa_{u} H_u \Phi^{\dagger}_u \sigma   \nonumber \\
& + \mathcal{Y}^{u}_{23} \overline{Q}_{2L} \widetilde{\Phi}_d u_{3R} +  \mathcal{Y}^{u}_{32} \overline{Q}_{3L} \widetilde{\Phi}_d u_{2R} + \kappa_{d} \widetilde{H}_d \Phi_d \sigma^{*}  \;.
\end{align}
The $ SM\times U(1)_{PQ} $ charges for the remaining fields are given by Table~\ref{tab:DFSZ1}.
Similarly, the down-quark sector can be written as
\begin{align}\label{eq:uv-4}
\mathcal{L}^{UV}_{d} &  \supset  \mathcal{Y}^{d}_{12} \overline{Q}_{1L} \Phi_d d_{2R} + \mathcal{Y}^{d}_{21} \overline{Q}_{2L} \Phi_d d_{1R} + \kappa_{u} H_d \Phi^{\dagger}_d \sigma   \nonumber \\
& + \mathcal{Y}^{d}_{23} \overline{Q}_{2L} \widetilde{\Phi}_d d_{3R} +  \mathcal{Y}^{d}_{32} \overline{Q}_{3L} \widetilde{\Phi}_d d_{2R} + \kappa_{d}\widetilde{H}_u \Phi_d \sigma^{*} \;.
\end{align}
The resulting quark mass matrices after the PQ and EW symmetry breaking can be read as
\begin{equation} \label{eq:m-ksvz-tii}
m_{u/d}  = \begin{pmatrix}
0 & \mathcal{Y}^{u/d}_{12} v_{\Phi_{u/d}}   & 0 \\ 
\mathcal{Y}^{u/d}_{21} v_{\Phi_{u/d}}   & 0 & \mathcal{Y}^{u/d}_{23} v_{\Phi_{d/u}}   \\ 
0 & \mathcal{Y}^{u/d}_{32}  v_{\Phi_{d/u}}  & y_{33}^{u/d} v_{u/d}
\end{pmatrix}\;, 
\end{equation} 
where the $ vev $s $v_{\Phi_{u/d}}$ of the additional doublets are determined by the scalar potential~\footnote{In~\cite{CentellesChulia:2020bnf},  the fermion mass hierarchies and the strong CP problem  with four Higgs doublets along with  the PQ symmetry have been discussed. }. We can write at leading order both vevs as \begin{equation}
v_{\Phi_{u/d}}\approx - \frac{\kappa_{u/d} v_\sigma v_{u/d}}{M^2_{\Phi_{u/d}}}
\end{equation} 
where $M_{\Phi_{u/d}}$ are heavy scalar masses. Finally, we like to remind here that the quark mass matrices in Eqs.~\ref{eq:massqf} and \ref{eq:m-ksvz-tii} have a NNI structure.

\begin{figure}
\centering
\includegraphics[width = 0.35\textwidth]{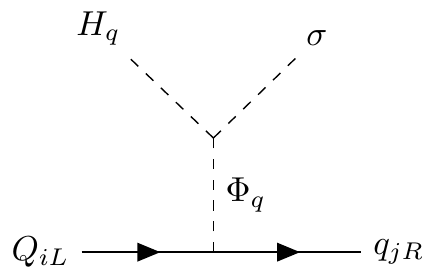}
\caption{\footnotesize UV complete diagram within the DFSZ type-II seesaw framework as apparent from Eqs. \ref{eq:uv-3}, and \ref{eq:uv-4}.}
\label{fig:uvdfsz2}
\end{figure}

\subsection{Lepton Sector}\label{sec:Lepton}
In our formalism, the Yukawa Lagrangian invariant under $SM\times U(1)_{PQ}$ for charged-leptons and neutrinos is given by
\begin{eqnarray}
-\mathcal{L}_{y}^l & \supset  y_{e}\overline{L}_{e L} H_d \ell_{eR} + y_{\mu}\overline{L}_{\mu L} \widetilde{H}_u \ell_{\mu R} + y_{\tau}\overline{L}_{\tau L} H_d\ell_{\tau R} \nonumber \\
& + y_{1}^{\nu}\overline{L}_{e L} H_u N_1 + y_{2}^{\nu}\overline{L}_{\mu L} \widetilde{H}_d N_2 + y_{3}^{\nu}\overline{L}_{\tau L} H_u N_3 \;.
\end{eqnarray}
%
The charge assignments for the leptonic fields are given in Table~\ref{tab:DFSZ1-Leptons}.

\begin{center}
\begin{table}[h] 
\scriptsize
\begin{tabular}{ c | c c c | c  }
\hline
Fields/Symmetry  & \hspace{1cm}  $L_{iL} $ \hspace{1cm} & \hspace{1cm}   $\ell_{iR} $ \hspace{1cm} &\hspace{1cm}  $N_{i}$ \hspace{1cm} & $\sigma^{\prime}$  \\
\hline
\hline
$SU(2)_L\times U(1)_Y$   &  (2, -1/2)  &  (1, -1) & (1, 0)  & (1, 0) \\  
\hline
$U(1)_{PQ}$  & (1, -3,  0) &  (0, -2,  -1) &  (0, -2, -1)  & 2 \\
\hline
  \end{tabular}
\caption{\footnotesize Field content and transformation properties of the leptonic fields and the scalar field $ \sigma^{\prime} $, where $i = 1, 2, 3$ represent the three lepton families.}\label{tab:DFSZ1-Leptons}
\end{table}
\end{center}

This leads to diagonal charged-lepton as well as Dirac neutrino mass matrices, respectively.
On the other hand, the Lagrangian involving right-handed neutrinos is given by~\footnote{Notice that the same leptonic Lagrangian was obtained in the context of a gauged $U(1)^\prime$ symmetry in~\cite{Flores:2020lji}.} 
\begin{eqnarray}
-\mathcal{L}_{Majorana} & \supset 
& M_1 \overline{N_1^c} N_1 + y_{12}^N \overline{N_1^c}N_2\sigma^{\prime} + y_{13}^N \overline{N_1^c}N_3 \sigma + y_{33}^N \overline{N_3^c}N_3 \sigma^{\prime}\;.
\end{eqnarray}
%
Therefore, the Majorana neutrino mass matrix takes the form 
\begin{equation}
M_R = \left( \begin{array}{ccc}
\times & \times & \times \\
\times & 0 & 0 \\
\times & 0 & \times
\end{array}\right)\; .
\end{equation}
Now, in the type-I seesaw formalism, a light neutrino mass matrix is given by $- m_{\nu} \approx M_D^T M_R^{-1} M_D  $, which can be read as
\begin{equation}
m_{\nu} = \left( \begin{array}{ccc}
0 & \times & 0 \\
\times & \times & \times \\
0 & \times & \times
\end{array}\right)\;,
\end{equation}
and it corresponds to the type $A_2$ neutrino mass matrix. 

It is to be noted  here that the inclusion of $ \sigma^{\prime} $ allows a non-zero dimension-6 terms in the quark sector, which are zero at dimension-5 as given by Eq.~\ref{eq:MassQuark}. The Yukawa Lagrangian at dimension-6  for the up-quark sector can be written as 
\begin{eqnarray}\label{ew:LagD6-up}
-\mathcal{L}^{d=6}_{Y} \supset  \frac{C^{u}_{13}}{\Lambda^2} \overline{Q}_{1L} H_u u_{3R} \sigma^{\prime 2} + \frac{C^{u}_{31}}{\Lambda^2} \overline{Q}_{3L} H_u u_{1R} \sigma^{\prime 2} + \frac{C^{u}_{22}}{\Lambda^2} \overline{Q}_{2L} \widetilde{H}_d u_{2R} \sigma^{\prime * 2}  \;.
\end{eqnarray}
However, we find that the UV-completion of the first term of Eq.~\ref{ew:LagD6-up} can only be achieved using the vector like fermions as given by Table~\ref{tab:DFSZ2}, whereas one needs new fermions to do the same for second and third terms, respectively. We explicitly show the UV-completion of the first term and the corresponding Lagrangian can be written as
\begin{align}\label{eq:uv-u6}
\mathcal{L}^{UV}_{u} &  \supset  \mathcal{Y}^{u}_{12} \overline{Q}_{1L} H_u F^{12}_{uR} + \mathcal{M}^{u}_{12} \overline{F^{12}_{uR}} F^{12}_{uL} + \mathcal{Y}^{u^\prime}_{12}~ \overline{F^{12}_{uL}} \sigma^{\prime} F^{32}_{uR} + \mathcal{M}^{u}_{32} \overline{F^{32}_{uR}} F^{32}_{uL} +  \mathcal{Y}^{u^\prime}_{32}~ \overline{F^{32}_{uL}} \sigma^{\prime} u_{3R} \;.
\end{align}

Similarly, the down-quark sector can be written as
\begin{eqnarray}\label{ew:LagD6-down}
-\mathcal{L}^{d=6}_{Y} \supset \frac{C^{d}_{13}}{\Lambda^2} \overline{Q}_{1L} H_d d_{3R} \sigma^{\prime 2} + \frac{C^{d}_{31}}{\Lambda^2} \overline{Q}_{3L} H_d d_{1R} \sigma^{\prime 2} + \frac{C^{d}_{22}}{\Lambda^2} \overline{Q}_{2L} \widetilde{H}_u d_{2R} \sigma^{\prime * 2}  \;.
\end{eqnarray}
Here, we also find that one can achieve the UV-completion only for the first term of Eq.~\ref{ew:LagD6-down} and the UV-complete Lagrangian can be written as
\begin{align}\label{eq:uv-d6}
\mathcal{L}^{UV}_{d} &  \supset  \mathcal{Y}^{d}_{12} \overline{Q}_{1L} H_d F^{12}_{dR} + \mathcal{M}^{d}_{12} \overline{F^{12}_{dR}} F^{12}_{dL} + \mathcal{Y}^{d^\prime}_{12}~ \overline{F^{12}_{dL}} \sigma^{\prime} F^{32}_{dR} + \mathcal{M}^{d}_{32} \overline{F^{32}_{dR}} F^{32}_{dL} +  \mathcal{Y}^{d^\prime}_{32}~ \overline{F^{32}_{dL}} \sigma^{\prime} d_{3R} \;.
\end{align}
%
\begin{figure}
\centering
\includegraphics[width=0.8\textwidth]{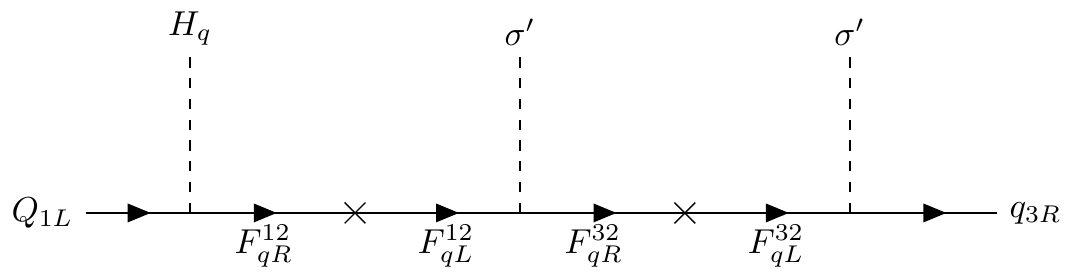}
\caption{\footnotesize UV complete diagram for a dimension-6 operator within the DFSZ type-I seesaw framework in presence of flavon field $ \sigma^\prime $ as follows from  Eqs.~\ref{eq:uv-u6}, and \ref{eq:uv-d6}.}
\label{fig:uvdiagramy}
\end{figure}

%
%

Therefore, in presence of dimension-6 terms, we generate a non-zero (1, 3) element of the quark mass matrix as given by Eq.~\ref{eq:MassQuark}.
However, in our analysis, we focus on the leading dimension-4 and 5 contributions to the mass matrices, noting that given the matter content proposed, this is the minimal set of operators needed to obtain the fermion masses and mixings compatible with observations. We assume higher order operators are either suppressed by a high enough energy scale to be subdominant, or forbidden by an additional symmetry.
\subsection{Scalar sectors}\label{sec:ScalarSector}
In minimal DFSZ models the $SU(2)_L$ singlet scalar field couples with the doublets either through the cubic $\widetilde{H}_u^\dagger H_d  \sigma$ term or through the quartic $ \widetilde{H}_u^\dagger H_d \sigma^2$ term \cite{Dine:1981rt,Espriu:2015mfa}. The $\widetilde{H}_u^\dagger H_d  \sigma$ term is PQ invariant if the PQ charge of $\sigma$ is chosen to be equal but opposite to the charge of the combination $\widetilde{H}_u^\dagger H_d$, while the $ \widetilde{H}_u^\dagger H_d \sigma^2$ term is PQ invariant if the PQ charge of $\sigma$ is chosen to be half the opposite charge of the combination $\widetilde{H}_u^\dagger H_d$. 
In this work, due to the charge convention chosen for the scalars, the couplings of $\widetilde{H}_u^\dagger H_d$ are to $\sigma^{*}$ instead of $\sigma$.
The PQ charges of the scalars are such that both the cubic, $\lambda_{10} \widetilde{H}^\dagger_uH_d \sigma^{\prime *}$, and quartic, $ \lambda_6  \widetilde{H}^\dagger_u H_d \sigma^{*2}$, couplings between the $SU(2)_L$ doublets and the respective singlets are present (see Appendix~\ref{AppendixPotential} for the complete scalar potential). We define the scalar fields present in the first DFSZ model as
\begin{equation}
H_u= \begin{pmatrix}
h_u^0 + i A_u \\ 
h_u^-
\end{pmatrix}\;,  \quad
H_d= \begin{pmatrix}
h_d^+  \\ 
h_d^0 + i A_d
\end{pmatrix}\;,  \quad
\sigma= S + i A \;,\quad
\sigma^\prime= S^\prime + i A^\prime \;,
\end{equation}
whereas the second DFSZ model contains two additional $SU(2)_L$ doublets $\Phi_u$ and $\Phi_d$, which we define as
\begin{equation}
\Phi_u= \begin{pmatrix}
\phi_u^0 + i A^\prime_u \\ 
\phi_u^-
\end{pmatrix}\;,  \quad
\Phi_d= \begin{pmatrix}
\phi_d^+  \\ 
\phi_d^0 + i A^\prime_d
\end{pmatrix}\;.
\end{equation}
The scalar potentials for both models have been provided in Appendix~\ref{AppendixPotential}.
For each model we may write the Goldstone mode eaten by the Z-boson as 
\begin{equation}
A_Z= \frac{\sum_i Y_i v_i A_i}{\sqrt{\sum_i{Y_i^2 v_i^2}}} \;,
\label{eq:ZGoldstone}
\end{equation}
where $Y_i$ is the hypercharge of each scalar field, $v_i$ its $vev$, for the type-I seesaw model $A_i\in \{A,A^\prime,A_u,A_d\}$ and $A_i\in \{A,A^\prime,A_u,A_d,A_u^\prime,A_d^\prime\}$ for the type-II seesaw. The Goldstone boson related to the PQ symmetry is likewise given by
\begin{equation}
A_{PQ}= \frac{\sum_i X_i v_i A_i}{\sqrt{\sum_i{X_i^2 v_i^2}}}\;,
\label{eq:PQGoldstone}
\end{equation}
where $X_i$ is the PQ charge of the scalar field to which $A_i$ belongs.
The physical axion, however, is orthogonal to the Z-boson Goldstone, so to obtain it from these two equations, we must perform the following subtraction~\cite{Srednicki:1985xd}
\begin{equation}
a=A_{PQ}- \left[\frac{\sum_i X_i Y_i v_i^2}{\sqrt{\sum_i{Y_i^2 v_i^2}}\sqrt{\sum_i{X_i^2 v_i^2}}} \right] A_Z\;.
\label{eq:physicalaxion}
\end{equation}
The possible values of $v_i$ for the doublet fields are bounded by the SM, $\sqrt{\sum_i{Y_i^2 v_i^2}} = 246~ \text{GeV}$. On the other hand, the possible values of $\sqrt{\sum_i{X_i^2 v_i^2}}$ are bounded from below by its relationship to the axion decay constant $f_a$. This means that there must be a strong hierarchy between the $vev$s of the $\sigma$ and $\sigma^\prime$ fields and the $vev$s of the doublet fields.
From Eqs.~{\ref{eq:ZGoldstone} - \ref{eq:physicalaxion} we can see that this results in the physical axion being composed to a high degree of only the $\sigma$ and $\sigma^\prime$ fields. Additionally, we choose to adopt the hierarchy $v_\sigma^\prime << v_\sigma$ as it leads to negligible mixing between the $A$ and $A^\prime$ fields, resulting in a stronger coupling of the axion to the quark fields, as in our framework quark masses are derived from the couplings of $\sigma$ at leading order. With these considerations in mind we extract the axion-quark couplings by setting $a\sim A$.
We would also like to mention that the details about the energy scale of $f_a$ and related phenomenology  are analyzed in subsequent sections. 
\subsection{Axion couplings and mass}\label{sec:AxionCoupling}
The couplings of the axion to the gluon and the photon are determined by the electromagnetic anomaly factor \textit{E} and the QCD anomaly factor \textit{N}. The effective Lagrangian that relates the axion couplings with  anomalies (\textit{E} and \textit{N}) can be read as
\begin{equation}
\mathcal{L} \supset \frac{\alpha_s}{8 \pi} \frac{a}{f_a} (F \widetilde{F})_C + \frac{E}{N} \frac{\alpha_{EM}}{8 \pi} \frac{a}{f_a} (F \widetilde{F})_{EM}\;,
\end{equation}
where $ (F \widetilde{F})_C = \frac{1}{2} \varepsilon^{\mu\nu\rho\sigma}F^b_{\mu\nu} F^b_{\rho\sigma} $, $ F^b_{\mu\nu} $ is the color field strength $(b = 1, . . . ,8 )$, and $  (F \widetilde{F})_{EM} = \frac{1}{2} \varepsilon^{\mu\nu\rho\sigma}F_{\mu\nu} F_{\rho\sigma}$, $ F_{\mu\nu} $ is the electromagnetic field strength.
The anomaly factors \textit{E} and \textit{N} are given by the expressions \cite{Srednicki:1985xd}
\begin{equation}
N=\sum\nolimits_{i} X_i T(R_i)\;, ~~ \quad E=\sum\nolimits_{i} X_i Q_i^2 D(R_i)\;,
\end{equation}
where the sums run over all fermions, $X_i$ is the PQ charge of fermion $i$, $T(R_i)$ is the index of the $SU(3)_c$ representation $R_i$ of fermion $i$ (in particular $ T(3) = 1/2,  T(6) = 5/2,  T(8) = 3 $).
Also,  $Q_i$ represent the  electric charge and $D(R_i)$ the dimension of $R_i$.
%
Moreover, the axion decay constant is given by~\cite{diCortona:2015ldu}
\begin{equation}
f_a=\frac{\sqrt{\sum_i X_i^2 v_i^2}}{\sqrt{2} N}\;,
\end{equation}
where this sum runs over all scalars with PQ charge $X_i$ and $ vev $ $v_i$. For simplicity we will consider the dominant contribution to $f_a$ be from $\sigma$, so we may write $f_a \sim 2 v_\sigma/N$. 
The axion mass in terms of the decay constant is \cite{diCortona:2015ldu}
\begin{equation}
m_a = 5.70 \, \mathrm{\mu eV} \left( \frac{10^{12} \, \text{GeV} }{f_a} \right)\;.
\end{equation}
Finally, the coefficients of the anomalous axion-gluon coupling and axion-photon coupling are given by~\cite{diCortona:2015ldu}
\begin{equation}\label{eq:Axion-Photon}
g_{a\gamma}= \frac{\alpha_{EM}}{2\pi f_a} \left(\frac{E}{N}-1.92 \right), \quad g_{ag}=\frac{\alpha_s}{8 \pi f_a}\;.
\end{equation}
In the Georgi-Glashow model~\cite{Georgi:1974sy} of gauge unification one can show that $E/N=8/3$, using this in Eq.~\ref{eq:Axion-Photon}, the axion-photon coupling can be expressed as 
\begin{equation}
g_{a\gamma}^{SU(5)}=\frac{1.53}{10^{16}\text{GeV}}\left(\frac{m_a}{\mu\text{eV} }\right)\;,
\end{equation}
where the Fine-structure constant $\alpha_{EM} = 1/137$ has been used. 
Here, for the DFSZ models the SM fermions are the only contributors to the anomalies, resulting in $N=5$ and $E=28/3$.
Therefore, compared to this benchmark the model presented here has the suppression
\begin{equation} \label{eq:gagamma}
\frac{|g_{a\gamma}^{SU(5)}|}{|g_{a\gamma}^{DFSZ}|} = 14\;. 
\end{equation}
For completeness, we briefly summarize here that the axion-photon interactions in the ($ g_{a\gamma} - m_a $) plane for $ m_a \geq 1$ eV is mostly constrained from  cosmology and astrophysics as can be seen from figure-1 of Ref.~\cite{Irastorza:2018dyq}.
Besides this,  for $ m_a \leq 1$  eV various haloscope detectors put the tightest constraints on the axion mass and axion-photon coupling as has been outlined in figure-16 of Ref.~\cite{DiLuzio:2020wdo}.
For the axion mass of $ \mathcal{O} (10^{-6}) $ eV, as the reason of interest for the model, it is the Axion Dark Matter eXperiment (ADMX)~\cite{Stern:2016bbw} searching for cold dark matter axions with a haloscope detector, provides the most stringent bound. It can be seen from figure-5 of~\cite{Stern:2016bbw} that the ADMX can explore $ 2\times 10^{-6} \leq m_a \leq  3.8\times 10^{-6}$ eV for the coupling strength of $ \mathcal{O} (10^{-15}) $ (GeV)$ ^{-1} $.
On the other hand, the ADMX Phase IIa/Gen-2 can improve their sensitivity of axion mass to  ($1.8,~ 8)\times 10^{-6}$ eV for $ |g_{a\gamma}|  $ one order smaller compared to the latest ADMX bound, i.e.,  $ \mathcal{O} (10^{-16}) $ (GeV)$ ^{-1} $, for details see figure-9 of~\cite{Stern:2016bbw}.
For definiteness, by choosing $f_a \sim 10^{12}$ GeV, and $m_a \sim  10^{-6}$ eV, we find $ |g_{a\gamma}| \sim  10^{-18} $ (GeV)$ ^{-1} $ using Eq.~\ref{eq:Axion-Photon} and hence
the suppression of this coupling places our model beyond the reach of projected ADMX Phase IIa/Gen-2 sensitivity \cite{Stern:2016bbw}.


We now proceed to discuss various low energy phenomenologies  within the FNPQ-model in subsequent sections. 
As per the quark sector goes, we consider the quark mass matrix as given by Eq.~\ref{eq:massqf} with a vanishing (1, 3) element and we adopt this as it can be seen from Eq.~\ref{eq:MassQuark} that the (1, 3) element arises from dimension-6 which is $1/\Lambda^2$ suppressed.
\section{Numerical Results}\label{sec:results}
\subsection{Masses and mixings of fermions}
Here we perform a $\chi^2$ analysis to find masses and mixing parameters of both quarks and leptons sectors.
First, a global $\chi^2$ fit is conducted to find the values of the parameters of the up and down quark matrices in our framework. The $\chi^2$ function takes the following form
\begin{equation}\label{eq:chisq}
\chi^2 = \sum \frac{(\mu_{exp} - \mu_{fit})^2}{\sigma_{exp}^2} \;,
\end{equation}
where the sum runs over all observables. Also, $\mu_{fit}$ represent the masses and mixings calculated using fitting parameters, as given by Eq.~\ref{eq:fitmassquark}, we list them in Table~\ref{tab:fit}.
Here $\mu_{exp}$ and $\sigma_{exp}$ are the observables and their standard deviation~\cite{Antusch:2013jca, Zyla:2020zbs} as given by  Table~\ref{tab:fit}. 

\begin{table}[t] 
\scriptsize
	\begin{minipage}{0.5 \linewidth}
	   	\centering
		\begin{tabular}{ c | c } 
		\hline
			Parameter   & \hspace{1cm}  Best fit \hspace{1cm}  \\
			\hline
			\hline
			$A_u/(10^{-2} \U{GeV})$ & $1.519$ \\
			$B_u/(10^{-2} \U{GeV})$ & $-5.368 $ \\
			$C_u/\U{GeV}$ & $-3.004 $ \\
			$D_u/(10^{1} \U{GeV})$ & $3.562 $ \\
			$E_u/(10^{2} \U{GeV)}$ & $1.679 $ \\
			$A_d/(10^{-2} \U{GeV})$ & $-1.233 $ \\
			$B_d/(10^{-2} \U{GeV})$ & $1.228$ \\
			$C_d/(10^{-1} \U{GeV})$ & $-3.074 $ \\
			$D_d/(10^{-1} \U{GeV})$ & $-4.782 $ \\
			$E_d/\U{GeV}$ & $-2.793 $ \\
			$\alpha_u/{}^{\circ}$ & $96.56$ \\
			$\beta_u/{}^{\circ}$ & $98.23$ \\ 
		\end{tabular}	
	\end{minipage}

	\begin{minipage}{0.5 \linewidth}
	   	\centering
	\begin{tabular}[t]{ l |c|c c |c }
\hline
\multirow{2}{*}{Observable}& \multicolumn{2}{c}{Global-fit value} & & \multirow{2}{*}{Model best-fit}  \\
		\cline{2-4}
		& Best-fit value & 1$\sigma$ range  &  & \\
			\hline
			\hline
			$\theta_{12}^q/{}^{\circ}$ & $13.09$ & $13.06 \to 13.12$  && $12.988$ \\
			$\theta_{13}^q/{}^{\circ}$ & $0.207$  & $0.202 \to 0.213$  && $0.2000$ \\
			$\theta_{23}^q/{}^{\circ}$ & $2.32$ & $ 2.29 \to 2.37$  && $2.381$ \\
			$\delta^q/{}^{\circ}$ & $68.53$ & $66.06 \to 71.10$   && $68.720$ \\
			$m_u/( 10^{-3} \U{GeV} )$ & $1.288$ & $0.766\to 1.550$  && $1.2743$ \\
			$m_c/(10^{-1} \U{GeV})$ & $6.268$ & $6.076 \to 6.459$  && $6.2592$ \\
			$m_t/ \U{GeV}$ & $171.68$ & $170.17 \to 173.18 $  && $171.687$ \\
			$m_d/( 10^{-3} \U{GeV})$ & $2.751$ & $2.577 \to 3.151$  && $2.7330$ \\
			$m_s/(10^{-2} \U{GeV}$) & $5.432$ & $5.153 \to 5.728$  && $5.4311$ \\
			$m_b/\U{GeV}$ & $2.854 $ & $2.827 \to 2.880$  && $2.8501$ \\
			$\chi_q^2$ &&&& 1.0901 \\ \hline
		\end{tabular}	
	\end{minipage}
\caption{\footnotesize Best-fit values of the model parameters in the quark sector are shown in the upper table. The global best-fit as well as their $ 1\sigma $ error \cite{Antusch:2013jca, Zyla:2020zbs} for the various observables  are given in the second and third columns of the lower table.   Also, the best-fit values of the various observables are listed in the last column of the lower table.}\label{tab:fit}
\end{table}

Note that by redefining the quark fields it can be shown that there exist only two non-zero phases. In the Appendix \ref{app:quark}, we give a detailed analysis of the phase redefinition. Thus, the up- and down-quark mass matrices that are used in the fit are given by
\begin{align}
m_{u/d} & =   \begin{pmatrix}
0  & A_{u/d}   & 0 \\
B_{u/d} \, e^{-i \alpha_{{u/d}}}  & 0 & C_{u/d} \, e^{-i \alpha_{{u/d}}} \\
0 & D_{u/d} \, e^{-i \beta_{{u/d}}}  & E_{u/d} \, e^{-i \beta_{{u/d}}} 
\end{pmatrix} \;.
\label{eq:fitmassquark}
\end{align}

Also, as pointed out in Appendix~\ref{app:quark}, it is the difference in the up- and down- quark phase matrices that are relevant for the quark mixing matrix, hence in our numerical analysis,  phases $ \alpha_d $ and $ \beta_d $ have been fixed to zero. There are 12 parameters that need to be fitted, 10 amplitudes (5 for each matrix) and 2 phases. 
We fit these 12 parameters to account for the 10 physical observables related to them, the 6 quark masses, the 3 CKM angles and the CP-violating phase.

The observables are obtained from these matrices using the MPT package \cite{Antusch:2005gp}. The fit is done at the energy scale $M_Z$~\cite{Antusch:2013jca}. 
The initial values for the fitting procedure are randomized.  The results from the best fit are given in Table~\ref{tab:fit}, where $\chi_q^2 = 1.0901$.

\begin{table}[!t] 
\scriptsize
	\begin{minipage}{0.5 \linewidth}
	   	\centering
		\begin{tabular}{ c | c } 
		\hline
			Parameter   & \hspace{1cm}  Best fit \hspace{1cm}  \\
			\hline
			\hline
			$a /(10^{-3} \U{eV})$ & $9.933$ \\
			$b /(10^{-2} \U{eV})$ & $2.646$ \\
			$c /(10^{-2} \U{eV})$ & $2.475 $ \\
			$d /(10^{-2} \U{eV})$ & $2.264$ \\
			$\phi_a/{}^{\circ}$ & $29.87$ \\
			$\phi_b/{}^{\circ}$ & $91.88$ \\
			$\phi_c/{}^{\circ}$ & $3.03$ \\
			$\phi_d /{}^{\circ}$ & $-109.97$ \\

		\end{tabular}	
	\end{minipage}

	\begin{minipage}{0.5 \linewidth}
	   	\centering
	\begin{tabular}[t]{ l |c|c c |c }
\hline
\multirow{2}{*}{Observable}& \multicolumn{2}{c}{Global-fit value} & & \multirow{2}{*}{Model best-fit}  \\
		\cline{2-4}
		& Best-fit value & 1$\sigma$ range  &  & \\
			\hline
			\hline
			$\theta_{12}^l/{}^{\circ}$ & $34.5$ & $33.5 \to 35.7$  && $34.85$ \\
			$\theta_{13}^l/{}^{\circ}$ & $8.45$  & $8.31 \to 8.61$  && $8.432$ \\
			$\theta_{23}^l/{}^{\circ}$ & $47.7$ & $46.0 \to 48.9$  && $48.11$ \\
			$\delta^l/{}^{\circ}$ & $218$ & $191 \to 256$   && $258.8$ \\
			$\alpha/{}^{\circ}$ &  &    && $65.27$ \\
			$\beta/{}^{\circ}$ &  &    && $265.08$ \\
			$\Delta m_{21}^2/( 10^{-5} \U{{eV}}^2 )$ & $7.55$  & $7.39 \to 7.75$  && $7.571$ \\
			$\Delta m_{32}^2/( 10^{-3} \U{{eV}}^2 )$ & $2.424$ & $2.394 \to 2.454$  && $2.4221$ \\
			$\sum m_\nu /(10^{-2} \U{eV})$ &  & && $6.453$\\
			$m_e/\U{MeV} $ & $0.4865763$  & $0.4865735 \to 0.4865789$  && - \\
			$m_\mu/ \U{GeV}$ & $0.10271897$ & $0.10271866 \to 0.10271931$  && - \\
			$m_\tau/ \U{GeV}$ & $1.74618$  & $1.74602 \to 1.74633$  && - \\

			$\chi_l^2$ &&&& 2.0053 \\ \hline
		\end{tabular}	
	\end{minipage}
\caption{\footnotesize Best-fit values of the model parameters in the lepton sector are shown in the upper table. The global best-fit as well as their $ 1\sigma $ error \cite{Antusch:2013jca, Zyla:2020zbs} for the various observables  are given in the second and third columns of the lower table.   Also, the best-fit values of the various observables are listed in the last column of the lower table.}\label{tab:fitlep}
\end{table}

Following a similar approach, we also find the masses and mixing angles of the leptonic sector. In this case only the elements of the neutrino mass matrix are fitted, since the charged lepton matrix is already diagonal. The neutrino mass matrix has the  $A_2$-texture and takes the form
\begin{align}
m_{\nu} & =   \begin{pmatrix}
0  & a \, e^{i\phi_a}  & 0 \\
a \, e^{i\phi_a}  & b \, e^{i\phi_b} & c \, e^{i\phi_c} \\
0 & c \, e^{i\phi_c}  & d \, e^{i\phi_d}
\end{pmatrix}.
\label{eq:fitmassneu}
\end{align}

The $\chi^2$ function is identical as in Eq.~\ref{eq:chisq}, the only differences being the observables and their standard deviations. The leptonic masses and mixings obtained from the fit, which are compatible with the latest global fit data, can be seen in Table~\ref{tab:fitlep} at $\chi_l^2 = 2.0053$. 

\subsection{Flavor Violating decays with axions}\label{sec:FVD-Axion}
The models described here contain flavor violating Yukawa couplings. Of particular interest are the axion flavor violating couplings to quarks  as they lead to the decays with final state axions $q_i\rightarrow q_j a$. These processes can be probed by meson decays with final state axions. These flavor violating couplings are common to both models presented. From the effective Lagrangian of Eqs. \ref{eq:type-1up} and \ref{eq:type-1down} we can write the Yukawa couplings to $\sigma$ in the interaction basis as 
\begin{equation}
\mathcal{L}_\sigma = y_{ij}^{u} \overline{u_{jL}} \sigma u_{iR}+ y_{ij}^{u^\prime} \overline{u_{jL}} \sigma^* u_{iR} + y_{ij}^{d} \overline{d_{jL}} \sigma d_{iR} +y_{ij}^{d^\prime} \overline{d_{jL}} \sigma^* d_{iR}  \;,
\end{equation}
where the Yukawa coupling matrices $y_{ij}^{u(^\prime)}$ and $y_{ij}^{d(^\prime)}$ are
\begin{equation}
y^{u/d} = \frac{1}{v_\sigma}
\begin{pmatrix}
0 & \mathbf{A}_{u/d} & 0 \\ 
\mathbf{B}_{u/d} & 0 & 0 \\ 
0 & 0 & 0
\end{pmatrix} \;,~~
y^{u^\prime/d^\prime} =\frac{1}{v_\sigma}
\begin{pmatrix}
0 & 0 & 0 \\ 
0 & 0 & \mathbf{C}_{u/d} \\ 
0 & \mathbf{D}_{u/d} & 0
\end{pmatrix}\;.
\end{equation}
In the quark mass basis the couplings to sigma are obtained by transforming the fermion fields to the physical basis
\begin{equation}
\mathcal{L}_{\sigma q^\prime}= y_{ij}^{u } U^\dagger_{ik L} \overline{u^\prime_{k L}} \sigma U_{jl R} u^\prime_{lR} + y_{ij}^{u ^\prime}  U^\dagger_{ik L} \overline{u^\prime_{k L}} \sigma^* U_{jl R} u^\prime_{lR} + y_{ij}^{d} V^\dagger_{ik L} \overline{d^\prime_{k L}} \sigma V_{jl R} d^\prime_{lR} + y_{ij}^{d ^\prime}  V^\dagger_{ik L} \overline{d^\prime_{k L}} \sigma^* V_{jl R} d^\prime_{lR}\;,
\label{eq:axionyukawas}
\end{equation}
where $u^\prime/d^\prime$ denote quark mass eigenstates, $U_L$ and $U_R$ diagonalize the up quark mass matrix and $V_L$ and $V_R$ diagonalize the down quark mass matrix, respectively.
By defining $\lambda^{u (^\prime)}= U^\dagger_L y^{u (^\prime)} U_R$ and $\lambda^{d (^\prime)}= V^\dagger_L y^{d (^\prime)} V_R$ we may write 
\begin{equation}
\mathcal{L}_{\sigma q^\prime}= \lambda_{ij}^{u } \overline{u^\prime_{i L}} \sigma u^\prime_{jR} + \lambda_{ij}^{u^\prime}  \overline{u^\prime_{i L}} \sigma^* u^\prime_{jR} + \lambda_{ij}^{d }  \overline{d^\prime_{i L}} \sigma d^\prime_{jR} + \lambda_{ij}^{d^\prime}  \overline{d^\prime_{i L}} \sigma^* d^\prime_{jR} \;.
\label{eq:physicalyukawas}
\end{equation}
Finally, by neglecting axion mixing with other scalars, we write the quark couplings to the axion as
\begin{equation}
\mathcal{L}_{a q^\prime}=i a ( \epsilon_{ij}^{u} \overline{u^\prime_{i}}  u^\prime_{j} +  \epsilon_{ij}^{d}  \overline{d^\prime_{i}}  d^\prime_{j}  + \epsilon_{ij}^{\prime u} \overline{u^\prime_{i}}\gamma_5  u^\prime_{j} +  \epsilon_{ij}^{\prime d}  \overline{d^\prime_{i}} \gamma_5 d^\prime_{j} ) \;,
\end{equation}
where $\epsilon_{ij}^{u,d}= (\lambda_{ij}-\lambda_{ij}^\dagger)/2 $ and $\epsilon_{ij}^{\prime u,d}= (\lambda_{ij}+\lambda_{ij}^\dagger)/2 $ \;.
From these couplings we may calculate the branching ratio of flavor violating decays with axions.
The most sensitive tests of neutral flavor violation with a final state axion is the $K^+ \rightarrow \pi^+ a$ process. The decay ratio for the Kaon decay to axion and pion is given by 
\begin{equation}
\Gamma(K^+\rightarrow\pi^+ a)\approx\frac{m_K}{64 \pi} |\epsilon_{21}^{d}|^2 B_S^2\left(1-\frac{m_\pi^2}{m_K^2}\right) \;,
\end{equation}
where $B_S$ is a non-perturbative parameter $B_S\sim 4.6$ \cite{Kamenik:2011vy}. To estimate the Kaon decay we use the best-fit point, as tabulated in Table~\ref{tab:fit},  in Eqs.~\ref{eq:axionyukawas} - \ref{eq:physicalyukawas}  to find $|\epsilon_{21}^{d }|^2$,  this leads to
\begin{equation}
\Gamma(K^+\rightarrow\pi^+ a)\approx \frac{ 1.9 \times 10^{-9} \text{GeV}^3}{v_\sigma^2}\;. 
\end{equation}
Using the latest constraint of the BR of  Kaon decay from the E949 Collaboration~\cite{Adler:2008zza} i.e., 
\begin{equation}
{\rm BR}(K^+\rightarrow\pi^+ a) = \frac{\Gamma(K^+\rightarrow\pi^+ a)}{\Gamma_{Total}(K^+)} < 7.3\times 10^{-11}\;,
\end{equation}
we may constrain $v_\sigma \geqslant 2.5 \times 10^{10} ~\text{GeV}$. 
For the models described above we have $v_\sigma \approx \sqrt{2} f_a N$ with $N=5$, and hence using this approximation one can translate bounds of $ v_\sigma $ to the axion decay constant $ f_a $, where
 \begin{equation}
f_a \geqslant 7 \times 10^9 \U{GeV}\;. 
\end{equation}
Analogously, we may consider the $B^+ \rightarrow K^+ a$ decay, where the bottom to strange quark transition is probed. The decay width of this process is given by
\begin{equation}
\Gamma(B^+\rightarrow K^+ a)\approx \frac{m_B}{64 \pi} |\epsilon_{32}^d|^2 (f_0^K(0))^2 \left(\frac{m_B}{m_b-m_s} \right)^2 \left(1-\frac{m_K^2}{m_B^2} \right)^3\;, 
\end{equation}
with $f_0^K(0)\sim 0.33$\cite{Ball:2004ye}.
The experiment Belle-II may constrain the branching ratio of this process to ${\rm BR}(B^+\rightarrow K^+ a)<10^{-6}-10^{-8}$ \cite{Calibbi:2016hwq}, which would lead to 
\begin{equation}
v_\sigma \geqslant 1.8 \times (10^7 - 10^8) ~\text{GeV}\;.
\end{equation} 
In our framework, $v_\sigma\approx \sqrt{2} N f_a$, the bound above translates to
\begin{equation}
f_a \geqslant 6 \times( 10^6 - 10^7)~ \text{GeV}.
\end{equation}
The axion decay constant $f_a > 7 \times 10^9$ GeV bound translates in our models to the equivalent limits $m_a < 0.7 \times 10^{-3}$ eV, and $|g_{a\gamma}(\text{GeV}^{-1})| < 0.8 \times 10^{-14}$. These bounds are two orders of magnitude stronger than the limits from astrophysics, see figure-1 of~\cite{Irastorza:2018dyq}, for completeness.
We end this section by noting that these bounds are obtained by neglecting $\sigma-\sigma^\prime$ mixing. When this mixing is sizable these bounds on $f_a$ are relaxed.
\subsection{Flavor Violating Higgs couplings}
%
The presence of two Higgs doublets in the DFSZ models raises the possibility of flavor changing neutral currents (FCNC) mediated by scalar particles. Moreover, the identification of the U(1) flavor symmetry with the Peccei-Quinn symmetry makes the implementation of natural flavor conservation difficult by requiring additional discrete symmetries and/or additional Higgs doublets. Therefore, in the DFSZ scheme of flavored axion models, scalar FCNCs are expected on minimal models. On the phenomenological side, these FCNCs are strongly constrained by processes, namely $K_L \rightarrow \mu^- \mu^+$ or top decays such as $t\rightarrow h c,h u$ \cite{Herrero-Garcia:2019mcy}. The masses of the additional scalar particles and the non-diagonal couplings of the light Higgs  may be constrained by the experimental limits on these processes. The high scale of PQ breaking induces a decoupling of the components of $H_u$ and $H_d$ from the components of $\sigma_i$. Hence, we assume one may write at leading order $h\approx h^u_0 \cos\alpha + h^d_0 \sin\alpha$, $H\approx -h^u_0 \sin\alpha + h^d_0 \cos\alpha$, where one identifies $h$ as the 125 GeV boson observed at LHC, and $H$ as an additional heavy scalar. The couplings of these two particles to SM fermions may be obtained from the effective Lagrangian and read as follows
\begin{equation}
\mathcal{L}\supset \frac{C^u_1}{v_u} \overline{u_L} u_R h_0^u + \frac{C^u_2}{v_d} \overline{u_L} u_R h_0^d + \frac{C^d_1}{v_d} \overline{d_L} d_R h_0^d + \frac{C^d_2}{v_u} \overline{d_L} d_R h_0^u  \;,
\end{equation}  
where the matrices $C^{u/d}_i$ are given by
\begin{equation} \label{eq:Ciu}
C^{u/d}_1= \begin{pmatrix}
0 & \mathbf{A}_{u/d} & 0 \\ 
\mathbf{B}_{u/d} & 0 & 0 \\ 
0 & 0 & \mathbf{E}_{u/d}
\end{pmatrix}\;,~~~
\mathbf{C}^{u/d}_2 = \begin{pmatrix}
0 & 0 & 0 \\ 
0 & 0 & \mathbf{C}_{u/d} \\ 
0 & \mathbf{D}_{u/d} & 0
\end{pmatrix}\; .
\end{equation}
In terms of the physical scalar and quark fields we have
\begin{equation}
\mathcal{L}\supset h \overline{u_L^{\prime}}u_R^\prime \left(\frac{C^{\prime u}_1}{v_u} \cos\alpha -\frac{C^{\prime u}_2}{v_d} \sin\alpha \right) + H \overline{u_L^{\prime}}u_R^\prime \left(\frac{C^{\prime u}_1}{v_u} \sin\alpha +\frac{C^{\prime u}_2}{v_d} \cos\alpha \right) + (u \rightarrow d) \;,
\label{eq:lighthiggsFVYukawas}
\end{equation}
where the matrices $C^{\prime u/d}_i$ are defined as
\begin{equation} \label{eq:Ciu2}
C^{\prime u}_i=U_L^\dagger C^{ u}_i U_R\quad ,\quad C^{\prime d}_i=V_L^\dagger C^{ d}_i V_R \;.
\end{equation} 
\begin{figure}[!t]
\includegraphics[scale=0.7]{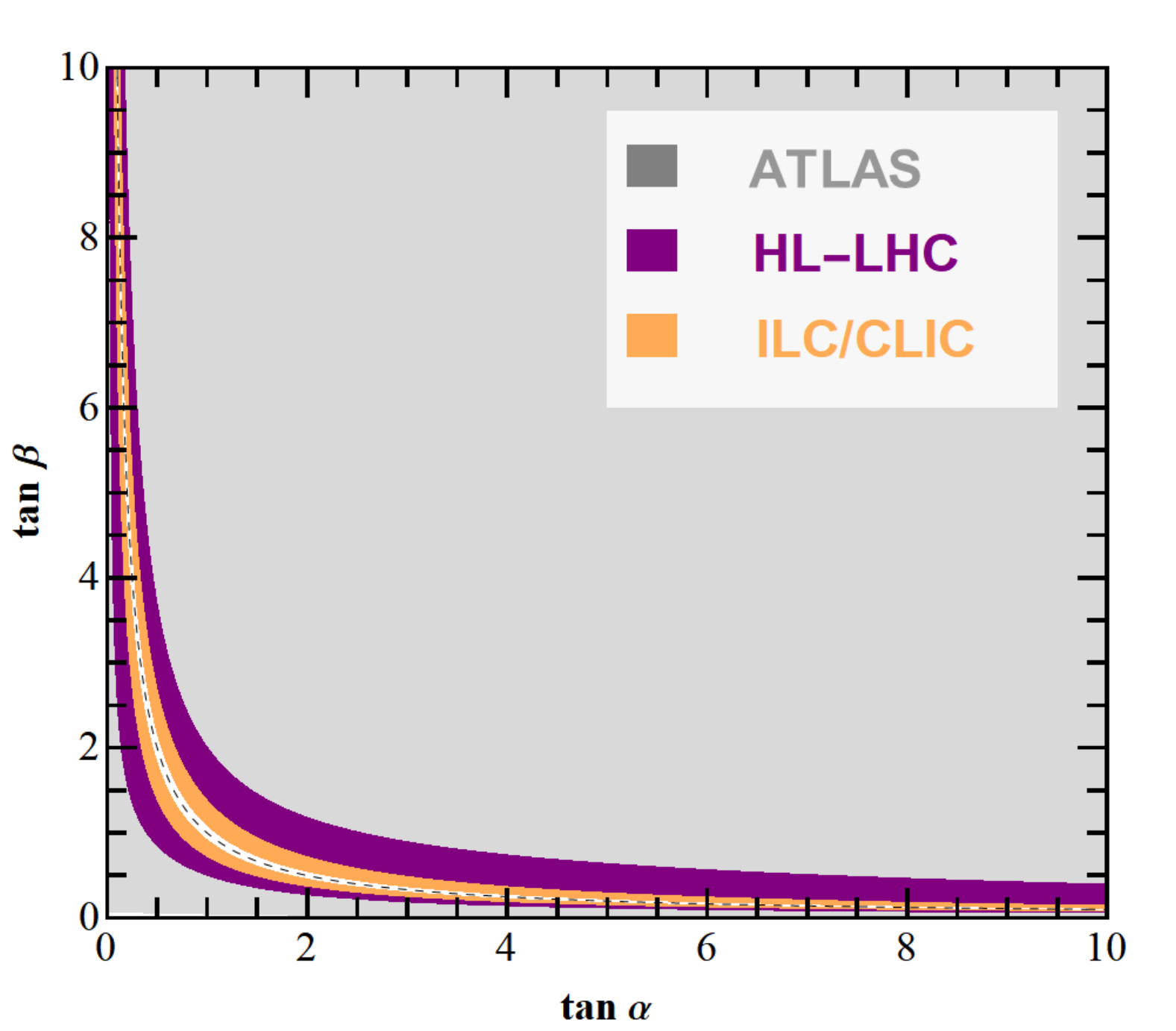}
\caption{\footnotesize Exclusion region plot in the $(\tan \alpha - \tan \beta)$ plane obtained from the non-observation of the $t \rightarrow h c$ flavor violating decay. The gray colored region is excluded by ATLAS data~\cite{Aaboud:2018oqm}, the purple colored region is expected to be probed in the future by the HL-LHC experiment \cite{ATLAS:2016qxw} and the orange region will be further probed by ILC or CLIC \cite{Vos:2016til}. The uncolored region (see white thin band) predicts a branching ratio beyond the sensitivity of these experiments. The dashed line indicates limit of no flavor violation in light Higgs Yukawa couplings. }
\label{fig:toptocharm}
\end{figure}
We see that in the limit $v_u/v_d =\cot\alpha$ the Yukawa couplings of $h$ are proportional to the up and down quark mass matrices, while the couplings of H are not. After the transformation of the fermion states to the mass basis the couplings of h are diagonal, while the couplings of H are not. This limit does not affect the couplings of the axion, which means that the processes discussed in the previous section do not vanish. A deviation of $\alpha$ from the flavor conserving limit will introduce flavor violating couplings of the light Higgs, which can be probed with observables such as $t\rightarrow h c$ in the up-sector and $h\rightarrow b s$ in the down sector. The $t\rightarrow hc$ decay channel currently has an upper bound set by ATLAS \cite{Aaboud:2018oqm}
\begin{equation}
{\rm BR}(t\rightarrow h c)_{\rm LHC}< 1.1 \times 10^{-3} \;,
\end{equation}
while future experiments HL-LHC~\cite{ATLAS:2016qxw}, ILC and CLIC~\cite{Vos:2016til} project the following sensitivities to this process
\begin{align}
{\rm BR}(t\rightarrow hc)_{\rm HL-LHC}&< 2 \times 10^{-4} \;,\\
{\rm BR}(t\rightarrow hc)_{\rm ILC/CLIC}&<  10^{-5} \nonumber\;.\\ 
\end{align}
Thus, the branching ratio can be calculated from Eq.~\ref{eq:lighthiggsFVYukawas} as \cite{Atwood:2013ica}
\begin{equation}
\Gamma_{t\rightarrow h c}= \frac{C_{tc}^2 m_t}{16\pi}\sqrt{\left[1-(R_c-R_h)^2 \right]\left[1-(R_c+R_h)^2\right]} \left[(R_c+1)^2-R_h^2\right] \;,
\label{eq:toptocharmBR}
\end{equation}
where the coupling $C_{tc}$ is defined as
\begin{equation}
 C_{tc}=\frac{\left[(C_{1}^{\prime u})_{23} +(C_{1}^{\prime u})_{32}\right]\cos\alpha }{v_{SM}\sin\beta}
 -\frac{\left[(C_{2}^{\prime u})_{23} +(C_{2}^{\prime u})_{32}\right]\sin\alpha}{v_{SM}\cos\beta} \;,
\end{equation}
$m_t$ is the top quark mass, $R_h$ is the higgs to top mass ratio $R_h=m_h/m_t$, and $R_c$ is the charm to top mass ratio $R_c = m_c/m_t$. Using the experimental value for the total width of the top quark \cite{Zyla:2020zbs} we derive the constraints on the free parameters $\tan \alpha$ and $\tan \beta$ as shown in Fig. \ref{fig:toptocharm}. We use the best fit point given in Table~\ref{tab:fit}, to obtain a numerical value of Eq.~\ref{eq:toptocharmBR} and we derive the approximate constraint
\begin{equation}
\left|\frac{\cos\alpha}{\sin\beta}\left(1- \tan\alpha \tan\beta  \right)\right| \leq  17 \frac{\Gamma_{t\rightarrow h c}^{Exp}}{\text[GeV]} \;,
\label{eq:numericaltoptocharmh}
\end{equation}
for a given experimental input of the decay width $\Gamma_{t\rightarrow h c}^{Exp}$.
We see from Fig.~\ref{fig:toptocharm} that, as expected from Eq.~\ref{eq:numericaltoptocharmh}, small values for $\tan \beta$ allow only large values of $\tan \alpha$ and vice versa. It can also be seen that ATLAS data has already ruled out a large portion of the parameter space (gray-region) and HL-LHC (purple-region) and CLIC (orange-region) will leave only a small region around the $\tan\beta=\cot\alpha$ limit unprobed (see Fig.~\ref{fig:toptocharm} caption for details).
Finally, we would like to mention that the $t\rightarrow h u$ and $h\rightarrow c u$ decays can also place constraints on $\alpha$ and $\beta$, however we find these to be numerically much weaker than the constraints from $t \rightarrow hc$, given that the hierarchy in the $C_i^{u}$ (see Eq.~\ref{eq:Ciu}) matrices is preserved in the physical $C_i^{\prime u}$ matrix (see Eq.~\ref{eq:Ciu2}).
%
Moreover, currently there are no experimental constraints on $h\rightarrow bs$ decays, although they are phenomenologically interesting in the context of two Higgs Doublet models \cite{Crivellin:2017upt}.
\subsection{Flavored Axion as Dark Matter candidate}
The axion can be a good dark matter candidate, provided a sufficient amount of them was produced in the early universe. There are several production mechanisms for axions.
The relic density produced by the misalignment mechanism is \cite{Fox:2004kb,Marsh:2015xka}
\begin{equation}\label{eq:DMRD}
\Omega_a h^2 \approx 2 \times 10^4 \left( \frac{f_a}{10^{16} \text{GeV}} \right)^{7/6} \langle \theta_{a, i}^2 \rangle \; , 
\end{equation}
where $\theta_{a, i}$ is the initial misalignment angle of the cosmological axion field and it lies in the range $(0, 2\pi)$. Now, we notice that for the axion breaking scale $ 5\times10^{10} < f_a <  1\times10^{15}$ (GeV), one can match the axion relic density to the observed dark matter relic abundance $ \Omega_{DM} h^{2} \sim 0.12$ for $ 0 < \theta_{a, i} < 2\pi $. It is worthwhile to mention that the $N>1$ prediction of DFSZ models induce the formation of stable domain walls in the universe. The existence of these stable domain walls is incompatible with the standard cosmology \cite{Sikivie:1982qv}. One way to avoid the effect of domain walls on the observed universe is to embed this type of models in a cosmological model where inflation happens after the formation of these walls, thereby inflating away the density of the walls. Another possible resolution of the domain wall problem is to destabilize the walls with a dynamical breaking of the PQ symmetry \cite{Barr:2014vva,Reig:2019vqh}.

\section{Conclusion}\label{sec:conclusion}
In this work, to address the hierarchal flavor structure of fermion masses as well as the strong CP problem  of QCD, we adopt a formalism based on the Froggatt-Nielsen symmetry. It is well known that the FN symmetry is one of the leading mechanisms to explain fermion masses and their mixings. 
On the other hand, the Peccei-Quinn symmetry provides an elegant solution to the strong CP problem.
%
Here, an attempt has been made by identifying the PQ symmetry $U(1)_{PQ}$ with the FN symmetry $U(1)_{FN}$ to address both the drawbacks of the SM.
In doing so, we assign PQ charges to the SM and BSM fields (see Table~\ref{tab:DFSZ1}), within the framework of the DFSZ style axion models,  in such a manner that  
the (3, 3) entry of the quark mass matrices are generated by a dimension-4 operator, whereas remaining entries are obtained at the dimension-5 level.
We end up with the Nearest-Neighbor-Interaction structure for the quark mass matrices in two different UV-completions, namely the type-I, and -II Dirac seesaw  mechanisms for quark masses}.  Moreover, upon the breaking of the assigned symmetry, a flavored-axion from the flavon fields has been obtained, which solves the strong CP problem and also is a Dark Matter candidate. For the lepton sector, we also extended the model by considering the type-I seesaw mechanism.
We obtained the $A_2$-type Majorana neutrino mass matrix in the diagonal charged lepton mass basis.

A comprehensive numerical analysis has been performed to find low energy fermion masses and mixings, which are summarized in Table~\ref{tab:fit}, \ref{tab:fitlep} and are consistent with the latest global-fit data. 
Besides this, based on the ratio between the electromagnetic to  the QCD anomaly factor, we have determined 
the axion-photon coupling, which is suppressed by a factor of 14 (as given by Eq.~\ref{eq:gagamma}) compared to the $ SU(5) $ GUT model.
The models discussed here also contain axion flavor violating couplings to quarks. Using these couplings, we have calculated the branching ratios of  $ K^+\rightarrow\pi^+ a $ and $ B^+\rightarrow K^+ a $ processes that involve the flavored-axion. These indicate that the axion decay constant is $f_a > 7 \times 10^9$ GeV, whereas the axion mass limit is $m_a < 0.7 \times 10^{-3}$ eV, and the axion-photon coupling is $|g_{a\gamma}(\text{GeV}^{-1})| < 0.8 \times 10^{-14}$.
We also have the possibility of flavor violating neutral currents mediated by scalar particles due to the presence of two Higgs doublets. The branching ratio for the $ t \rightarrow hc $ decay channel has been calculated for the model and the results are summarized in Fig.~\ref{fig:toptocharm}, where the latest bounds of ATLAS as well as the  future experiments limits of HL-LHC, ILC and CLIC
have been incorporated.
Finally, we have pointed out  that when the axion breaking scale lies in the range $ f_a \in (5\times10^{10},   1\times10^{15})$ GeV, then the  axion relic density can be matched to the observed dark matter relic abundance.  
\acknowledgements
This work is supported by 
the grants CONACYT CB-2017-2018/A1-S-13051 (M\'exico), DGAPA-PAPIIT IN107118, DGAPA-PAPIIT IN107621 and SNI (M\'exico). NN is supported by the postdoctoral fellowship program DGAPA-UNAM. LMGDLV is supported by CONACYT.

\appendix
\section{Phase redefinition of quark mass matrices}\label{app:quark}
The up- and down-quark mass matrices  as given by Eq.~\ref{eq:massqf} can be written as

\begin{equation}\label{eq:qmassA1}
m_{u/d} = \begin{pmatrix}
0  & A_{u/d} \, e^{i \alpha_{u/d}} & 0 \\
B_{u/d} \, e^{i \beta_{u/d}} & 0 & C_{u/d} \, e^{i \gamma_{u/d}} \\
0 & D_{u/d} \, e^{i \delta_{u/d}} & E_{u/d} \, e^{i \epsilon_{u/d}} \\
\end{pmatrix} \;,
\end{equation}
with $A, \, B, \, C, \, D, \, E, \, \alpha$ and $ \beta $ as real parameters. 
The above mass matrix  $ m_{u/d} $ can be diagonalized by bi-unitary transformation of the form
\begin{equation}\label{eq:qmassA2}
m^{diag} = V^{\dagger}_L m V_R = O^T P^\dagger_L m P_R O\;,
\end{equation}
 where $ L $ and $R$ depict the left- and right-chiral fields, respectively. Also, $V_L = P_L O$ and $V_R = P_R O$ are the unitary matrices that diagonalize $ m^\dagger m $  and $mm^\dagger$ and $P_L =  \mathrm{diag}(1, e^{i \alpha},  e^{ i \beta})$, $P_R = \mathrm{diag}(e^{i \rho_1}, e^{i \rho_2},  e^{i \rho_3})$ are the diagonal phase matrices, respectively. Notice that we drop subscript $({u/d})$ to demonstrate phase redefinition as the following formalism is same for both the up and down sector.
Now, given the form of $ P_L $ and $P_R$, one can construct a real solution of the  quark mass matrix (Eq.~\ref{eq:qmassA1}) following the transformation $ P^\dagger_L m P_R $ as mentioned by Eq.~\ref{eq:qmassA2}. Thus, the  quark mass matrix in terms of real parameters can be written as
\begin{equation}\label{eq:qmassA3}
m =  \begin{pmatrix}
0  & A \, & 0 \\
B \, & 0 & C \, \\
0 & D \,  & E \, \\
\end{pmatrix} \;.
\end{equation}
Notice that for most of the application $ V_R $ is irrelevant, it is the $ V_L $ that enters in the CKM parameterization. Therefore, one can write quark mixing matrix as
\begin{align}
 V_{CKM} & = (V^{\dagger}_L)_u (V_L)_d =  (O^T P^\dagger_L)_u (P_L O)_d \;, \nonumber \\
 & = O^T_u  \mathrm{diag}(1, e^{-i (\alpha_u - \alpha_d)},  e^{ - i (\beta_u - \beta_d)}) O_d  \;.  
\end{align}
We see here that it is the difference in the diagonal-phase matrix that enters in the $ V_{CKM} $. Thus, in our numerical an analysis we have used the  quark mass matrix  of the form
\begin{equation}\label{eq:qmassA4}
 P^\dagger_L m = \begin{pmatrix}
0  & A \,  & 0 \\
B \, e^{- i \alpha}  & 0 & C \, e^{- i \alpha} \\
0 & D \,e^{- i \beta} & E \, e^{- i \beta} \\
\end{pmatrix} \;.
\end{equation}

\section{Scalar potentials}
\label{AppendixPotential}
Here, we give a full scalar potential for both the type-I and -II DFSZ seesaw models corresponding to their charge assignments as given by Tables~\ref{tab:DFSZ1}, \ref{tab:DFSZ1-Leptons}, respectively. For the type-I DFSZ seesaw model the full scalar potential is given by
\begin{eqnarray}
V &=& \mu_{u}^2 H_u^\dagger H_u + \mu_{d}^2 H_d^\dagger H_d + \mu_1^2 \sigma \sigma^* + \mu_2^2 \sigma^\prime\sigma^{\prime *} + \lambda_u (H_u^\dagger H_u)^2 + \lambda_d (H_d^\dagger H_d)^2 + \lambda ( \sigma \sigma^*)^2 + \lambda^\prime (\sigma^\prime\sigma^{\prime *})^2  \nonumber \\
  &+& \lambda_1  (H_u^\dagger H_d)(H_d^\dagger H_u) + \lambda_2  (H_u^\dagger H_u)(H_d^\dagger H_d) + \lambda_3   (H_u^\dagger  H_d)(H_d^\dagger  H_u) +  \lambda_4 (\sigma \sigma^*)(H_u^\dagger H_u)\nonumber \\
  & + & \lambda_5 (\sigma \sigma^*)(H_d^\dagger H_d) + \lambda_6  (\widetilde{H}^\dagger_u H_d)(\sigma^*)^2 
+ \lambda_7 (\sigma^\prime\sigma^{\prime *})(H_u^\dagger H_u) +  \lambda_8 (\sigma^\prime\sigma^{\prime*})(H_d^\dagger H_d) + \lambda_9 (\sigma^\prime\sigma^{\prime *})(\sigma \sigma^*) \nonumber \\
& + & \kappa (\sigma^2 \sigma^{\prime *}+(\sigma^*)^2 \sigma^{\prime} ) 
 + \lambda_{10} (\widetilde{H}^\dagger_uH_d)\sigma^{\prime *} \;. 
  \label{eq:potential1}
\end{eqnarray}
For the type-II DFSZ seesaw model we enlarge the previous model by two additional Higgs Doublets transforming as $\Phi_u \sim (-1/2,2)$ and $\Phi_d \sim (+1/2,2)$ under $(U(1)_Y,U(1)_{PQ})$. Therefore the scalar potential contains the following terms, in addition to those in \ref{eq:potential1}:
\begin{eqnarray}
V 	&=& \mu_{\Phi_u}^2 \Phi_u^\dagger \Phi_u  + \mu_{\Phi_d}^2 \Phi_d^\dagger \Phi_d  + \kappa_u (H_u^\dagger \Phi_u) \sigma^* + \kappa_d (H_d^\dagger \Phi_d) \sigma^* + \lambda_{11} (\widetilde{\Phi}_d^\dagger \Phi_u) (\sigma^{\prime *})^2 + \lambda_{12} (\widetilde{H}_d^\dagger \Phi_u) (\sigma^{\prime *} \sigma^* \nonumber)  \\
 &+& \lambda_{13} (\widetilde{H}_u^\dagger \Phi_d)  (\sigma^{\prime *} \sigma^*) + \lambda_{14} (\Phi_u^\dagger \Phi_u)^2 + \lambda_{15} (\Phi_d^\dagger \Phi_d)^2 + \lambda_{16} (\Phi_d^\dagger H_d)(H_d^\dagger \Phi_d) +  \lambda_{17} (\Phi_u^\dagger H_d)(H_d^\dagger \Phi_u) \nonumber  \\
 &+& \lambda_{18} (\Phi_d^\dagger H_u)(H_u^\dagger \Phi_d) 	+ \lambda_{19} ( \Phi_u^\dagger H_u)(H_u^\dagger \Phi_u) + \lambda_{20}  ( \Phi_d^\dagger \Phi_u) (\Phi_u^\dagger \Phi_d) + \lambda_{21} (\sigma \sigma^*)(\Phi_u^\dagger \Phi_u) \nonumber \\
 &+&  \lambda_{22} (\sigma \sigma^*)(\Phi_d^\dagger \Phi_d) + \lambda_{23} (\sigma^\prime\sigma^{\prime *})(\Phi_u^\dagger \Phi_u) + \lambda_{24} (\sigma^\prime\sigma^{\prime*})(\Phi_d^\dagger \Phi_d) 
	\;.
\label{eq:potential2}
\end{eqnarray}
\bibliographystyle{utphys}
\bibliography{bibliography}
\end{document}